\documentstyle{article}
\begin{document}

\immediate\write16{<WARNING: FEYNMAN macros work only with emTeX-dvivers
                    (dviscr.exe, dvihplj.exe, dvidot.exe, etc.) >}
\newdimen\Lengthunit
\newcount\Nhalfperiods
\Lengthunit = 1.5cm
\Nhalfperiods = 9
\catcode`\*=11
\newdimen\L*   \newdimen\d*   \newdimen\d**
\newdimen\dm*  \newdimen\dd*  \newdimen\dt*
\newdimen\a*   \newdimen\b*   \newdimen\c*
\newdimen\a**  \newdimen\b**
\newdimen\xL*  \newdimen\yL*
\newcount\k*   \newcount\l*   \newcount\m*
\newcount\n*   \newcount\dn*  \newcount\r*
\newcount\N*   \newcount\*one \newcount\*two  \*one=1 \*two=2
\newcount\*ths \*ths=1000
\def\GRAPH(hsize=#1)#2{\hbox to #1\Lengthunit{#2\hss}}
\def\Linewidth#1{\special{em:linewidth #1}}
\Linewidth{.4pt}
\def\sm*{\special{em:moveto}}
\def\sl*{\special{em:lineto}}
\newbox\spm*   \newbox\spl*
\setbox\spm*\hbox{\sm*}
\setbox\spl*\hbox{\sl*}
\def\mov#1(#2,#3)#4{\rlap{\L*=#1\Lengthunit\kern#2\L*\raise#3\L*\hbox{#4}}}
\def\smov#1(#2,#3)#4{\rlap{\L*=#1\Lengthunit
\xL*=\xscale\L*\yL*=\yscale\L*\kern#2\xL*\raise#3\yL*\hbox{#4}}}
\def\mov*(#1,#2)#3{\rlap{\kern#1\raise#2\hbox{#3}}}
\def\lin#1(#2,#3){\rlap{\sm*\mov#1(#2,#3){\sl*}}}
\def\arr*(#1,#2,#3){\mov*(#1\dd*,#1\dt*){%
\sm*\mov*(#2\dd*,#2\dt*){\mov*(#3\dt*,-#3\dd*){\sl*}}%
\sm*\mov*(#2\dd*,#2\dt*){\mov*(-#3\dt*,#3\dd*){\sl*}}}}
\def\arrow#1(#2,#3){\rlap{\lin#1(#2,#3)\mov#1(#2,#3){%
\d**=-.012\Lengthunit\dd*=#2\d**\dt*=#3\d**%
\arr*(1,10,4)\arr*(3,8,4)\arr*(4.8,4.2,3)}}}
\def\arrlin#1(#2,#3){\rlap{\L*=#1\Lengthunit\L*=.5\L*%
\lin#1(#2,#3)\mov*(#2\L*,#3\L*){\arrow.1(#2,#3)}}}
\def\dasharrow#1(#2,#3){\rlap{%
{\Lengthunit=0.9\Lengthunit\dashlin#1(#2,#3)\mov#1(#2,#3){\sm*}}%
\mov#1(#2,#3){\sl*\d**=-.012\Lengthunit\dd*=#2\d**\dt*=#3\d**%
\arr*(1,10,4)\arr*(3,8,4)\arr*(4.8,4.2,3)}}}
\def\clap#1{\hbox to 0pt{\hss #1\hss}}
\def\ind(#1,#2)#3{\rlap{%
\d*=.1\Lengthunit\kern#1\d*\raise#2\d*\hbox{\lower2pt\clap{$#3$}}}}
\def\sh*(#1,#2)#3{\rlap{%
\dm*=\the\n*\d**\xL*=\xscale\dm*\yL*=\yscale\dm*
\kern#1\xL*\raise#2\yL*\hbox{#3}}}
\def\calcnum*#1(#2,#3){\a*=1000sp\b*=1000sp\a*=#2\a*\b*=#3\b*%
\ifdim\a*<0pt\a*-\a*\fi\ifdim\b*<0pt\b*-\b*\fi%
\ifdim\a*>\b*\c*=.96\a*\advance\c*.4\b*%
\else\c*=.96\b*\advance\c*.4\a*\fi%
\k*\a*\multiply\k*\k*\l*\b*\multiply\l*\l*%
\m*\k*\advance\m*\l*\n*\c*\r*\n*\multiply\n*\n*%
\dn*\m*\advance\dn*-\n*\divide\dn*2\divide\dn*\r*%
\advance\r*\dn*%
\c*=\the\Nhalfperiods5sp\c*=#1\c*\ifdim\c*<0pt\c*-\c*\fi%
\multiply\c*\r*\N*\c*\divide\N*10000}
\def\dashlin#1(#2,#3){\rlap{\calcnum*#1(#2,#3)%
\d**=#1\Lengthunit\ifdim\d**<0pt\d**-\d**\fi%
\divide\N*2\multiply\N*2\advance\N*1%
\divide\d**\N*\sm*\n*\*one\sh*(#2,#3){\sl*}%
\loop\advance\n*\*one\sh*(#2,#3){\sm*}\advance\n*\*one\sh*(#2,#3){\sl*}%
\ifnum\n*<\N*\repeat}}
\def\dashdotlin#1(#2,#3){\rlap{\calcnum*#1(#2,#3)%
\d**=#1\Lengthunit\ifdim\d**<0pt\d**-\d**\fi%
\divide\N*2\multiply\N*2\advance\N*1\multiply\N*2%
\divide\d**\N*\sm*\n*\*two\sh*(#2,#3){\sl*}\loop%
\advance\n*\*one\sh*(#2,#3){\kern-1.48pt\lower.5pt\hbox{\rm.}}%
\advance\n*\*one\sh*(#2,#3){\sm*}%
\advance\n*\*two\sh*(#2,#3){\sl*}\ifnum\n*<\N*\repeat}}
\def\shl*(#1,#2)#3{\kern#1#3\lower#2#3\hbox{\unhcopy\spl*}}
\def\trianglin#1(#2,#3){\rlap{\toks0={#2}\toks1={#3}\calcnum*#1(#2,#3)%
\dd*=.57\Lengthunit\dd*=#1\dd*\divide\dd*\N*%
\d**=#1\Lengthunit\ifdim\d**<0pt\d**-\d**\fi%
\multiply\N*2\divide\d**\N*\advance\N*-1\sm*\n*\*one\loop%
\shl**{\dd*}\dd*-\dd*\advance\n*2%
\ifnum\n*<\N*\repeat\n*\N*\advance\n*1\shl**{0pt}}}
\def\wavelin#1(#2,#3){\rlap{\toks0={#2}\toks1={#3}\calcnum*#1(#2,#3)%
\dd*=.23\Lengthunit\dd*=#1\dd*\divide\dd*\N*%
\d**=#1\Lengthunit\ifdim\d**<0pt\d**-\d**\fi%
\multiply\N*4\divide\d**\N*\sm*\n*\*one\loop%
\shl**{\dd*}\dt*=1.3\dd*\advance\n*1%
\shl**{\dt*}\advance\n*\*one%
\shl**{\dd*}\advance\n*\*two%
\dd*-\dd*\ifnum\n*<\N*\repeat\n*\N*\shl**{0pt}}}
\def\w*lin(#1,#2){\rlap{\toks0={#1}\toks1={#2}\d**=\Lengthunit\dd*=-.12\d**%
\N*8\divide\d**\N*\sm*\n*\*one\loop%
\shl**{\dd*}\dt*=1.3\dd*\advance\n*\*one%
\shl**{\dt*}\advance\n*\*one%
\shl**{\dd*}\advance\n*\*one%
\shl**{0pt}\dd*-\dd*\advance\n*1\ifnum\n*<\N*\repeat}}
\def\l*arc(#1,#2)[#3][#4]{\rlap{\toks0={#1}\toks1={#2}\d**=\Lengthunit%
\dd*=#3.037\d**\dd*=#4\dd*\dt*=#3.049\d**\dt*=#4\dt*\ifdim\d**>16mm%
\d**=.25\d**\n*\*one\shl**{-\dd*}\n*\*two\shl**{-\dt*}\n*3\relax%
\shl**{-\dd*}\n*4\relax\shl**{0pt}\else\ifdim\d**>5mm%
\d**=.5\d**\n*\*one\shl**{-\dt*}\n*\*two\shl**{0pt}%
\else\n*\*one\shl**{0pt}\fi\fi}}
\def\d*arc(#1,#2)[#3][#4]{\rlap{\toks0={#1}\toks1={#2}\d**=\Lengthunit%
\dd*=#3.037\d**\dd*=#4\dd*\d**=.25\d**\sm*\n*\*one\shl**{-\dd*}%
\n*3\relax\sh*(#1,#2){\xL*=\xscale\dd*\yL*=\yscale\dd*
\kern#2\xL*\lower#1\yL*\hbox{\sm*}}%
\n*4\relax\shl**{0pt}}}
\def\arc#1[#2][#3]{\rlap{\Lengthunit=#1\Lengthunit%
\sm*\l*arc(#2.1914,#3.0381)[#2][#3]%
\smov(#2.1914,#3.0381){\l*arc(#2.1622,#3.1084)[#2][#3]}%
\smov(#2.3536,#3.1465){\l*arc(#2.1084,#3.1622)[#2][#3]}%
\smov(#2.4619,#3.3086){\l*arc(#2.0381,#3.1914)[#2][#3]}}}
\def\dasharc#1[#2][#3]{\rlap{\Lengthunit=#1\Lengthunit%
\d*arc(#2.1914,#3.0381)[#2][#3]%
\smov(#2.1914,#3.0381){\d*arc(#2.1622,#3.1084)[#2][#3]}%
\smov(#2.3536,#3.1465){\d*arc(#2.1084,#3.1622)[#2][#3]}%
\smov(#2.4619,#3.3086){\d*arc(#2.0381,#3.1914)[#2][#3]}}}
\def\wavearc#1[#2][#3]{\rlap{\Lengthunit=#1\Lengthunit%
\w*lin(#2.1914,#3.0381)%
\smov(#2.1914,#3.0381){\w*lin(#2.1622,#3.1084)}%
\smov(#2.3536,#3.1465){\w*lin(#2.1084,#3.1622)}%
\smov(#2.4619,#3.3086){\w*lin(#2.0381,#3.1914)}}}
\def\shl**#1{\c*=\the\n*\d**\d*=#1%
\a*=\the\toks0\c*\b*=\the\toks1\d*\advance\a*-\b*%
\b*=\the\toks1\c*\d*=\the\toks0\d*\advance\b*\d*%
\a*=\xscale\a*\b*=\yscale\b*%
\raise\b*\rlap{\kern\a*\unhcopy\spl*}}
\def\wlin*#1(#2,#3)[#4]{\rlap{\toks0={#2}\toks1={#3}%
\c*=#1\l*\c*\c*=.01\Lengthunit\m*\c*\divide\l*\m*%
\c*=\the\Nhalfperiods5sp\multiply\c*\l*\N*\c*\divide\N*\*ths%
\divide\N*2\multiply\N*2\advance\N*1%
\dd*=.002\Lengthunit\dd*=#4\dd*\multiply\dd*\l*\divide\dd*\N*%
\d**=#1\multiply\N*4\divide\d**\N*\sm*\n*\*one\loop%
\shl**{\dd*}\dt*=1.3\dd*\advance\n*\*one%
\shl**{\dt*}\advance\n*\*one%
\shl**{\dd*}\advance\n*\*two%
\dd*-\dd*\ifnum\n*<\N*\repeat\n*\N*\shl**{0pt}}}
\def\wavebox#1{\setbox0\hbox{#1}%
\a*=\wd0\advance\a*14pt\b*=\ht0\advance\b*\dp0\advance\b*14pt%
\hbox{\kern9pt%
\mov*(0pt,\ht0){\mov*(-7pt,7pt){\wlin*\a*(1,0)[+]\wlin*\b*(0,-1)[-]}}%
\mov*(\wd0,-\dp0){\mov*(7pt,-7pt){\wlin*\a*(-1,0)[+]\wlin*\b*(0,1)[-]}}%
\box0\kern9pt}}
\def\rectangle#1(#2,#3){%
\lin#1(#2,0)\lin#1(0,#3)\mov#1(0,#3){\lin#1(#2,0)}\mov#1(#2,0){\lin#1(0,#3)}}
\def\dashrectangle#1(#2,#3){\dashlin#1(#2,0)\dashlin#1(0,#3)%
\mov#1(0,#3){\dashlin#1(#2,0)}\mov#1(#2,0){\dashlin#1(0,#3)}}
\def\waverectangle#1(#2,#3){\L*=#1\Lengthunit\a*=#2\L*\b*=#3\L*%
\ifdim\a*<0pt\a*-\a*\def\x*{-1}\else\def\x*{1}\fi%
\ifdim\b*<0pt\b*-\b*\def\y*{-1}\else\def\y*{1}\fi%
\wlin*\a*(\x*,0)[-]\wlin*\b*(0,\y*)[+]%
\mov#1(0,#3){\wlin*\a*(\x*,0)[+]}\mov#1(#2,0){\wlin*\b*(0,\y*)[-]}}
\def\calcparab*{%
\ifnum\n*>\m*\k*\N*\advance\k*-\n*\else\k*\n*\fi%
\a*=\the\k* sp\a*=10\a*\b*\dm*\advance\b*-\a*\k*\b*%
\a*=\the\*ths\b*\divide\a*\l*\multiply\a*\k*%
\divide\a*\l*\k*\*ths\r*\a*\advance\k*-\r*%
\dt*=\the\k*\L*}
\def\arcto#1(#2,#3)[#4]{\rlap{\toks0={#2}\toks1={#3}\calcnum*#1(#2,#3)%
\dm*=135sp\dm*=#1\dm*\d**=#1\Lengthunit\ifdim\dm*<0pt\dm*-\dm*\fi%
\multiply\dm*\r*\a*=.3\dm*\a*=#4\a*\ifdim\a*<0pt\a*-\a*\fi%
\advance\dm*\a*\N*\dm*\divide\N*10000%
\divide\N*2\multiply\N*2\advance\N*1%
\L*=-.25\d**\L*=#4\L*\divide\d**\N*\divide\L*\*ths%
\m*\N*\divide\m*2\dm*=\the\m*5sp\l*\dm*%
\sm*\n*\*one\loop\calcparab*\shl**{-\dt*}%
\advance\n*1\ifnum\n*<\N*\repeat}}
\def\arrarcto#1(#2,#3)[#4]{\L*=#1\Lengthunit\L*=.54\L*%
\arcto#1(#2,#3)[#4]\mov*(#2\L*,#3\L*){\d*=.457\L*\d*=#4\d*\d**-\d*%
\mov*(#3\d**,#2\d*){\arrow.02(#2,#3)}}}
\def\dasharcto#1(#2,#3)[#4]{\rlap{\toks0={#2}\toks1={#3}\calcnum*#1(#2,#3)%
\dm*=\the\N*5sp\a*=.3\dm*\a*=#4\a*\ifdim\a*<0pt\a*-\a*\fi%
\advance\dm*\a*\N*\dm*%
\divide\N*20\multiply\N*2\advance\N*1\d**=#1\Lengthunit%
\L*=-.25\d**\L*=#4\L*\divide\d**\N*\divide\L*\*ths%
\m*\N*\divide\m*2\dm*=\the\m*5sp\l*\dm*%
\sm*\n*\*one\loop%
\calcparab*\shl**{-\dt*}\advance\n*1%
\ifnum\n*>\N*\else\calcparab*%
\sh*(#2,#3){\kern#3\dt*\lower#2\dt*\hbox{\sm*}}\fi%
\advance\n*1\ifnum\n*<\N*\repeat}}
\def\*shl*#1{%
\c*=\the\n*\d**\advance\c*#1\a**\d*\dt*\advance\d*#1\b**%
\a*=\the\toks0\c*\b*=\the\toks1\d*\advance\a*-\b*%
\b*=\the\toks1\c*\d*=\the\toks0\d*\advance\b*\d*%
\raise\b*\rlap{\kern\a*\unhcopy\spl*}}
\def\calcnormal*#1{%
\b**=10000sp\a**\b**\k*\n*\advance\k*-\m*%
\multiply\a**\k*\divide\a**\m*\a**=#1\a**\ifdim\a**<0pt\a**-\a**\fi%
\ifdim\a**>\b**\d*=.96\a**\advance\d*.4\b**%
\else\d*=.96\b**\advance\d*.4\a**\fi%
\d*=.01\d*\r*\d*\divide\a**\r*\divide\b**\r*%
\ifnum\k*<0\a**-\a**\fi\d*=#1\d*\ifdim\d*<0pt\b**-\b**\fi%
\k*\a**\a**=\the\k*\dd*\k*\b**\b**=\the\k*\dd*}
\def\wavearcto#1(#2,#3)[#4]{\rlap{\toks0={#2}\toks1={#3}\calcnum*#1(#2,#3)%
\c*=\the\N*5sp\a*=.4\c*\a*=#4\a*\ifdim\a*<0pt\a*-\a*\fi%
\advance\c*\a*\N*\c*\divide\N*20\multiply\N*2\advance\N*-1\multiply\N*4%
\d**=#1\Lengthunit\dd*=.012\d**\ifdim\d**<0pt\d**-\d**\fi\L*=.25\d**%
\divide\d**\N*\divide\dd*\N*\L*=#4\L*\divide\L*\*ths%
\m*\N*\divide\m*2\dm*=\the\m*0sp\l*\dm*%
\sm*\n*\*one\loop\calcnormal*{#4}\calcparab*%
\*shl*{1}\advance\n*\*one\calcparab*%
\*shl*{1.3}\advance\n*\*one\calcparab*%
\*shl*{1}\advance\n*2%
\dd*-\dd*\ifnum\n*<\N*\repeat\n*\N*\shl**{0pt}}}
\def\triangarcto#1(#2,#3)[#4]{\rlap{\toks0={#2}\toks1={#3}\calcnum*#1(#2,#3)%
\c*=\the\N*5sp\a*=.4\c*\a*=#4\a*\ifdim\a*<0pt\a*-\a*\fi%
\advance\c*\a*\N*\c*\divide\N*20\multiply\N*2\advance\N*-1\multiply\N*2%
\d**=#1\Lengthunit\dd*=.012\d**\ifdim\d**<0pt\d**-\d**\fi\L*=.25\d**%
\divide\d**\N*\divide\dd*\N*\L*=#4\L*\divide\L*\*ths%
\m*\N*\divide\m*2\dm*=\the\m*0sp\l*\dm*%
\sm*\n*\*one\loop\calcnormal*{#4}\calcparab*%
\*shl*{1}\advance\n*2%
\dd*-\dd*\ifnum\n*<\N*\repeat\n*\N*\shl**{0pt}}}
\def\hr*#1{\clap{\xL*=\xscale\Lengthunit\vrule width#1\xL* height.1pt}}
\def\shade#1[#2]{\rlap{\Lengthunit=#1\Lengthunit%
\smov(0,#2.05){\hr*{.994}}\smov(0,#2.1){\hr*{.980}}%
\smov(0,#2.15){\hr*{.953}}\smov(0,#2.2){\hr*{.916}}%
\smov(0,#2.25){\hr*{.867}}\smov(0,#2.3){\hr*{.798}}%
\smov(0,#2.35){\hr*{.715}}\smov(0,#2.4){\hr*{.603}}%
\smov(0,#2.45){\hr*{.435}}}}
\def\dshade#1[#2]{\rlap{%
\Lengthunit=#1\Lengthunit\if#2-\def\t*{+}\else\def\t*{-}\fi%
\smov(0,\t*.025){%
\smov(0,#2.05){\hr*{.995}}\smov(0,#2.1){\hr*{.988}}%
\smov(0,#2.15){\hr*{.969}}\smov(0,#2.2){\hr*{.937}}%
\smov(0,#2.25){\hr*{.893}}\smov(0,#2.3){\hr*{.836}}%
\smov(0,#2.35){\hr*{.760}}\smov(0,#2.4){\hr*{.662}}%
\smov(0,#2.45){\hr*{.531}}\smov(0,#2.5){\hr*{.320}}}}}
\def\vdot{\rlap{\kern-1.9pt\lower1.8pt\hbox{$\scriptstyle\bullet$}}}
\def\vtimes{\rlap{\kern-3pt\lower1.8pt\hbox{$\scriptstyle\times$}}}
\def\vDot{\rlap{\kern-2.3pt\lower2.7pt\hbox{$\bullet$}}}
\def\vTimes{\rlap{\kern-3.6pt\lower2.4pt\hbox{$\times$}}}
\catcode`\*=12
\newcount\CatcodeOfAtSign
\CatcodeOfAtSign=\the\catcode`\@
\catcode`\@=11
\newcount\n@ast
\def\n@ast@#1{\n@ast0\relax\get@ast@#1\end}
\def\get@ast@#1{\ifx#1\end\let\next\relax\else%
\ifx#1*\advance\n@ast1\fi\let\next\get@ast@\fi\next}
\newif\if@up \newif\if@dwn
\def\up@down@#1{\@upfalse\@dwnfalse%
\if#1u\@uptrue\fi\if#1U\@uptrue\fi\if#1+\@uptrue\fi%
\if#1d\@dwntrue\fi\if#1D\@dwntrue\fi\if#1-\@dwntrue\fi}
\def\halfcirc#1(#2)[#3]{{\Lengthunit=#2\Lengthunit\up@down@{#3}%
\if@up\smov(0,.5){\arc[-][-]\arc[+][-]}\fi%
\if@dwn\smov(0,-.5){\arc[-][+]\arc[+][+]}\fi%
\def\lft{\smov(0,.5){\arc[-][-]}\smov(0,-.5){\arc[-][+]}}%
\def\rght{\smov(0,.5){\arc[+][-]}\smov(0,-.5){\arc[+][+]}}%
\if#3l\lft\fi\if#3L\lft\fi\if#3r\rght\fi\if#3R\rght\fi%
\n@ast@{#1}%
\ifnum\n@ast>0\if@up\shade[+]\fi\if@dwn\shade[-]\fi\fi%
\ifnum\n@ast>1\if@up\dshade[+]\fi\if@dwn\dshade[-]\fi\fi}}
\def\halfdashcirc(#1)[#2]{{\Lengthunit=#1\Lengthunit\up@down@{#2}%
\if@up\smov(0,.5){\dasharc[-][-]\dasharc[+][-]}\fi%
\if@dwn\smov(0,-.5){\dasharc[-][+]\dasharc[+][+]}\fi%
\def\lft{\smov(0,.5){\dasharc[-][-]}\smov(0,-.5){\dasharc[-][+]}}%
\def\rght{\smov(0,.5){\dasharc[+][-]}\smov(0,-.5){\dasharc[+][+]}}%
\if#2l\lft\fi\if#2L\lft\fi\if#2r\rght\fi\if#2R\rght\fi}}
\def\halfwavecirc(#1)[#2]{{\Lengthunit=#1\Lengthunit\up@down@{#2}%
\if@up\smov(0,.5){\wavearc[-][-]\wavearc[+][-]}\fi%
\if@dwn\smov(0,-.5){\wavearc[-][+]\wavearc[+][+]}\fi%
\def\lft{\smov(0,.5){\wavearc[-][-]}\smov(0,-.5){\wavearc[-][+]}}%
\def\rght{\smov(0,.5){\wavearc[+][-]}\smov(0,-.5){\wavearc[+][+]}}%
\if#2l\lft\fi\if#2L\lft\fi\if#2r\rght\fi\if#2R\rght\fi}}
\def\Circle#1(#2){\halfcirc#1(#2)[u]\halfcirc#1(#2)[d]\n@ast@{#1}%
\ifnum\n@ast>0\clap{%
\dimen0=\xscale\Lengthunit\vrule width#2\dimen0 height.1pt}\fi}
\def\wavecirc(#1){\halfwavecirc(#1)[u]\halfwavecirc(#1)[d]}
\def\dashcirc(#1){\halfdashcirc(#1)[u]\halfdashcirc(#1)[d]}
%
\def\xscale{1}
\def\yscale{1}
\def\Ellipse#1(#2)[#3,#4]{\def\xscale{#3}\def\yscale{#4}%
\Circle#1(#2)\def\xscale{1}\def\yscale{1}}
\def\dashEllipse(#1)[#2,#3]{\def\xscale{#2}\def\yscale{#3}%
\dashcirc(#1)\def\xscale{1}\def\yscale{1}}
\def\waveEllipse(#1)[#2,#3]{\def\xscale{#2}\def\yscale{#3}%
\wavecirc(#1)\def\xscale{1}\def\yscale{1}}
\def\halfEllipse#1(#2)[#3][#4,#5]{\def\xscale{#4}\def\yscale{#5}%
\halfcirc#1(#2)[#3]\def\xscale{1}\def\yscale{1}}
\def\halfdashEllipse(#1)[#2][#3,#4]{\def\xscale{#3}\def\yscale{#4}%
\halfdashcirc(#1)[#2]\def\xscale{1}\def\yscale{1}}
\def\halfwaveEllipse(#1)[#2][#3,#4]{\def\xscale{#3}\def\yscale{#4}%
\halfwavecirc(#1)[#2]\def\xscale{1}\def\yscale{1}}
\catcode`\@=\the\CatcodeOfAtSign

~\hfill hep-th/9802155

\vspace*{0.5cm}

\begin{center}
{\bf \large EFFECTIVE ACTION OF $N=2$ SUPERSYMMETRIC \\ FIELD
THEORIES IN HARMONIC SUPERSPACE APPROACH}\\

\vspace*{0.5cm}

I.L. Buchbinder\\
Department of Theoretical Physics\\
Tomsk State Pedagogical University\\
634041, Tomsk, Russia
\end{center}

\begin{abstract}
The paper is a brief review of
the works devoted to problem of effective action in
$N=2$ super Yang-Mills theories in harmonic superspace
approach. The formulation of $N=2$ superfield models
in harmonic superspace is discussed, background field
method for $N=2$ super Yang-Mills theory is constructed
and general structure of effective action in harmonic
superspace is investigated. It is shown how the holomorphic
and non-holomorphic contributions to effective action
can be calculated within the harmonic superspace approach.
\end{abstract}

\section{Introduction}

$N=2$ supersymmetric field theories possess remarkable properties both at
the classical and quantum levels. Applications of $N=2$ supersymmetry in a
whole are very enormous and range from superstring theory to supergauge
field models and
so  called special geometry (see [1] for a modern review).
 In particular, $N=2$ super Yang-Mills theories are
finite beyond one-loop approximation [2-7]. A modern interest to quantum aspects
of $N=2$ supersymmetric field theory was inspired by the work of Seiberg and
Witten [8] where non-pertubative contribution to low-energy effective action was
exactly found. The Seiberg-Witten approach has essentialy been based on the
structure of perturbative low-energy effective action proposed by Seiberg [9].

The paper under consideration is devoted to discussing a systematic method
of investigation of effective action in
$N=2$ supersymmetric field theories within the harmonic superspace approach.
This approach has been developed by A.Galperin, E.Ivanov, S.Kalitzin,
V.Ogievetsky, E. Sokatchev  [10, 11] and it is very well adapted to be applied for studing
quantum aspects of $N=2$ supersymmetric field theories. From point of view of
quantum field theory the harmonic superspace approach possesses by two
essential  dignities:

(i) The $N=2$ supersymmetric field theories are formulated in this approach in
terms of $N=2$ superfields defined on suitable superspace.

(ii) The theories are defined by actions depending on unconstrained superfields.

The first feature garantees the manifest $N=2$ supersymmetry on all stages of
consideration. The second one allows to use the standart notions of
propagators and vertices and hence to apply a standart quantum field theory
technique.

The aspects of effective action problem for $N=2$, supersymmetric field theories
have been investigated in recent papers [12]. However all calculations were given
in these papers using a formulation of $N=2$ theories in terms of $N=1$ superfields
where $N=2$ supersymmetry is not manifest. As well known although such a formulation
is correct and sometimes convenient it does not allow to control the calculations
on a base of explicit symmetry and can
 lead to different kinds of miraculous cancellations an
actual reason of which is not seen. From this point of view developing a completely
$N=2$ supersymmetric approach to effective action problem looks like useful and
important enough.

This paper is a brief review of such an approach  which has been recently developed
in the refs.[13, 14] and allows  to preserve manifest $N=2$ supersymmetry at all steps
of effective action calculation.

\section{Brief Review of Harmonic Superspace Approach}

The aim of harmonic superspace approach is to formulate $N=2$ supersymmetric field
theories in terms of $N=2$ unconstrained superfields. A basic idea is to introduce a
superspace preserving $N=2$ supersymmetry but having lesser number of anticommuting
coordinates then in case of general $N=2$ superspace.

A first step leading to suitable superspace is based on introducing  the new bosonic
coordinates $u^{+}_{i}, u^{-}_{i}; i=1, 2$ forming a matrix belonging to $SU(2)$-group. These
coordinates $u^{\pm}_i$ are called the harmonics. Using the decomposition
\begin{eqnarray}
\theta^{\pm}_{\bar \alpha}= u^{\pm}_i \theta^i_{\alpha}, \qquad
\bar \theta^{\pm}_{\dot \alpha}= u^{\pm}_i \bar \theta^i_{\dot \alpha},
\end{eqnarray}
where $\theta^i_{\alpha}, \theta^i_{\dot \alpha}$ are the anticommuting
 coordinates of general $N=2$ superspace we introduce
the new space-time coordinates
\begin{eqnarray}
x^m_A = x^m - i(\theta^i \sigma^m \bar \theta^j +
\theta^j \sigma^m \bar \theta^i)u^+_i u^-_j
\end{eqnarray}
and get a superspace wich the coordinates
\begin{eqnarray}
(x^m_{A}, \theta^{\pm}_{\alpha}, \bar \theta^{\pm}_{\dot \alpha}, u^{\pm}_i).
\end{eqnarray}

A second step begins with decomposition of spinor derivatives
\begin{eqnarray}
D^{\pm}_{\alpha}=u^{\pm}_i D^i_{\alpha}, \qquad
\bar D^{\pm}_{\dot \alpha}=u^{\pm}_i \bar D^i_{\dot \alpha}
\end{eqnarray}
where $D^i_{\alpha}, \bar D^i_{\dot \alpha}$ are the supercovariant spinor derivatives
in general $N=2$ superspace
with the coordinates $(x^m, \theta^{i}_{\alpha}, \bar \theta^{i}_{\dot \alpha})$.
One can show that the derivatives $D^+_{\alpha}, \bar D^+_{\dot \alpha}$ have the very
simple form
\begin{eqnarray}
D^+_{\alpha}= \frac{\partial}{\partial \theta^{-}_{\alpha}}, \qquad
\bar D^+_{\dot \alpha}= \frac{\partial}{\partial \bar \theta^{-}_{\dot \alpha}}
\end{eqnarray}
Taking these derivatives one can consider the superfields
$\Phi(x^m_{A}, \theta^{\pm}_{\alpha}, \bar \theta^{\pm}_{\dot \alpha}, u^{\pm}_i)$
 satisfying the constraints
\begin{eqnarray}
D^+_{\alpha} \Phi = 0, \quad
\bar D^+_{\dot \alpha} \Phi = 0
\end{eqnarray}
The constraints (6) show that we  have a possibility to work with
 superfields
$\Phi = \Phi(x^m_{A}, \theta^{+}_{\alpha}, \bar \theta^{+}_{\dot \alpha}, u^{\pm}_i)$
 which do not depend on
$\theta^-_{\alpha}, \theta^-_{\dot \alpha}$ but preserve manifest $N=2$ supersymmetry.
Let us introduce a superspace with the coordinates
$(x^m_{A}, \theta^{+}_{\alpha}, \bar \theta^{+}_{\dot \alpha}, u^{\pm}_i)$.
It is remarkable that these
coordinates transform through each other under $N=2$ supersymmetry transformations
with full $N=2$ transformation parameters. It alloys to treat the above
superspace as an independent object. This superspace is called analytic subspace and
it plays in $N=2$ supersymmetry the same role as chiral subspace in $N=1$
supersymmetry. The superfields defined on analytic subspace are called analytic.
It is evident that any analytic superfields contains the same number of anticommuting
coordinates as general $N=1$ superfield. It leads to reducing a number of independent
components in compare with general $N=2$ superfields. However all components
depend now on extra bosonic coordinates $u^{\pm}_i$. Therefore any analytic superfield
contains infinite number of component fields from point of view of conventional
supersymmetric field theory. It is worth to point out that a possibility to construct the analytic
subspace is main dignity of harmonic superspace approach.

A third step is a construction of the action describing a dynamics of the superfields. Let
us start with matter superfields which are called the hypermultiplets. A simplest
hypermultiplet is described by analytic superfield
$q^+ (x^m_A, \theta^{+}_{\alpha}, \bar \theta^{+}_{\dot \alpha}, u^{\pm}_i)$
with $U(1)$-charge +1. The action of the free theory looks as follows
\begin{eqnarray}
S[\breve{q}^+,q^+] = - \int d \zeta^{(-4)} du \breve{q}^+ D^{++} q^+
\end{eqnarray}
where $d\zeta^{(4)}=d^4 x_A d^2 \theta^+ d^2 \bar \theta^+ $. Here
$\breve{q}^+$ means a special conjugation.
Integral over harmonics $d u$ was defined in the
papers [10, 11] and
\begin{eqnarray}
D^{++}=u^{+i} \frac{\partial}{\partial u^{-i}} -
2i(\theta^+ \sigma^m \bar \theta^+ )\frac{\partial}{\partial x^{m}_A} +
 \theta^{+ \alpha} \frac{\partial}{\partial \theta^{- \alpha}}+
\bar \theta^{+ \dot \alpha} \frac{\partial}{\partial \bar \theta^{- \dot \alpha}}
\end{eqnarray}
is a specific operator acting on harmonics.

Another hypermultiplet is described by real analytic superfield
$\omega (x_{A}, \theta^{+}, \bar \theta^{+}, u^{\pm})$ and has the action
\begin{eqnarray}
S[ \omega] = \int d \zeta^{(-4)} du (D^{++}
\omega)(D^{++} \omega)
\end{eqnarray}
Both $q^+$ and $\omega$ are the
unconstrained superfields.

The equations of motion for hypermultiplets look like as follows
\begin{eqnarray}
& &D^{++}q^+ = 0
\nonumber
\\
& &(D^{++})^2 \omega = 0
\end{eqnarray}
If we expand the $q^+$ and $\omega$ in a power series in harmonics and substitute
into equations of motion
we will see that only a few first terms of expansion are not equal to zero on-shell.
Hence, almost all $u$-dependence of $q^+$ and $\omega$ becomes to be unessential
on-shell and the
problem of extra bosonic coordinates dissapears.

To construct interacting theory one introduces the covariant derivatives
\begin{eqnarray}
\nabla^{++} = D^{++} + iV^{++}
\end{eqnarray}
where $V^{++} = V^{++ a}T^a , V^{++ a}$ is  analytic superfields with $U(1)$-charge +2
and $T^a$ are the generators
of internal symmetry. Namely this superfield $V^{++ a}$ is an unconstrained prepotential
for $N=2$ super Yang-Mills theory.

The final step is action for $V^{++}$. It can be written in the form [15]
\begin{eqnarray}
S_{SYM}[V^{++}] = \frac{1}{g^2} \int d^4 x d^8 \theta
\sum_{n=2}^{\infty} \frac{(-i)^n}{n}
\int du_1 ... du_n \frac{tr V^{++}(z, u_1)...V^{++}(z, u_n)}
{(u^+_1 u^+_2)...(u^+_n u^+_1)}
\end{eqnarray}
Here $z=(x^m,\theta^i_{\alpha},\bar \theta^i_{\dot \alpha})$,
$(u^+_1 u^+_2)=u^{+ i}_1 u^+_{2 i}$ and $g$ is  a coupling. This action is invariant
under the gauge transformations [10, 11]
\begin{eqnarray}
\delta V^{++} = - D^{++}\Lambda + ...
\end{eqnarray}
where $\Lambda$ is analytic superfield parameter. Taking into account this gauge
 transformation
one can  impose specific gauge fixing condition where practically  whole $u$-dependence
becomes to be unessential.

As result we obtain a formulation of super Yang-Mills theory coupled to a matter in a
manifest $N=2$ supersymmetric form in terms of unconstrained superfields.

\section{Background Field Method for $N=2$ Super Yang-Mills Theories}

Background field method is a some specific construction of effective action in gauge
field theories allowing to preserve classical gauge invariance in quantum theory. The
matter is, to quantize a gauge theory we impose the gauge fixing conditions and destroy
the classical gauge invariance of the theory. As a result, the effective action generally
speaking is not invariant under initial gauge transformations. Only S-matrix is
gauge independent object. However, there is a way to construct effective action which
will be invariant under the same gauge transformations as the initial classical action.
This way is based on splitting the fields into two pieces, the classical fields and the quantum
ones and imposing the gauge fixing conditions only on quantum fields. In concrete theories
one can find the suitable gauge fixing functions allowing to preserve classical
gauge invariance with respect to above classical fields which are the functional arguments
of the effective action.

Let us split the superfield $V^{++}$ into background $V^{++}$ and
quantum ${v}^{++}$ pisces
\begin{eqnarray} V^{++} \rightarrow V^{++} +
g v^{++} \end{eqnarray} where $V^{++}$ is background field and
$v^{++}$ is quantum one. Then the initial gauge transformations can be
realized in two different ways

\noindent
(i) background transformations
\begin{eqnarray}
\delta V^{++} &=& - D^{++}\Lambda - i[V^{++}, \Lambda] = -\nabla^{++}\Lambda
\nonumber
\\
\delta {v}^{++} &=& i [\Lambda, v^{++}]
\end{eqnarray}

\noindent
(ii) quantum transformations
\begin{eqnarray}
\delta V^{++} &=& 0
\nonumber
\\
\delta v^{++} &=& - \frac{1}{g}\nabla^{++}\Lambda - i[
v^{++}, \Lambda]
\end{eqnarray}

It is worth to point out here that the form of background - quantum splitting and corresponding
background and quantum transformations are absolutely analogous to the conventional
Yang-Mills theory but not to $N=1$ super Yang-Mills theory. Our aim is to construct effective action
as a functional of background field,  invariant under background gauge transformations.

After substituting the background -quantum splitting into the action we obtain
\begin{eqnarray}
& &S_{SYM}[V^{++}+ g{v}^{++}] = S_{SYM}[V^{++}]+
\nonumber
\\
&+& \frac{1}{4g} tr \int d\zeta^{(-4)}du v^{++}\bar D_{\dot \alpha}^+
\bar D^{ + \dot \alpha} \bar W_{(\lambda)} +
\Delta S_{SYM}[V^{++}, v^{++}]
\nonumber
\\
& &\Delta S_{SYM}[V^{++}, v^{++}] = - tr \int d^{12}z
 \sum^{\infty}_{n=2} \frac{(-ig)^{n-2}}{n} \int du_1 ... du_n
\frac{v^{++}_{(\tau)}(z,u_1)v^{++}_{(\tau)}(z, u_2) ...
v^{++}_{(\tau)}(z, u_n)}
{(u^+_1 u^+_2)(u^+_2 u^+_3) ... (u^+_n u^+_1)}
\nonumber
\\
& &\bar W_{(\lambda)} = e^{i\Omega} \bar W e^{-i\Omega}
\nonumber
\\
& &v^{++}_{(\tau)} = e^{-i\Omega} v^{++} e^{i\Omega}
\end{eqnarray}
The superfield $\Omega$ is called the bridge and it has been introduced in ref. [10]. The superfields $W$ and $\bar W$ are
the $N=2$ strengths introduced in ref. [17]. In the case under consideration  the $\Omega$
corresponds to
background superfield $V^{++}$. The action $\Delta S_{SYM}$ depends on $V^{++}$ via dependence
 $v^{++}_{(\tau)}$ on $V^{++}$. We want to point
out that each term in $\Delta S_{SYM}$ is manifestly invariant under background gauge transformations.

To quantize a theory within background field method we should fix only quantum gauge transformations.
One introduces the gauge fixing functions in the form
\begin{eqnarray}
{\cal F}^{(4)}_{(\tau)} = D^{++} v^{++}_{(\tau)}=
e^{-i\Omega}(\nabla^{++}v^{++}) e^{i\Omega} =
 e^{-i\Omega}{\cal F}^{(4)} e^{i\Omega}
\end{eqnarray}
and applies Faddeev-Popov procedure. It leads to the following form for effective action
$\Gamma_{SYM} [V^{++}]$
\begin{eqnarray}
e^{i\Gamma_{SYM}[V^{++}]} = e^{iS_{SYM}[V^{++}]}\int {\cal D}v^{++}
{\cal D}b {\cal D}c e^{i(\Delta S_{SYM}[v^{++}, V^{++}] +
S_{FP})} \delta ({\cal F}^{(4)} - f^{(4)})
\end{eqnarray}
Here $S_{FP}$
is Faddeev-Popov ghost action
 \begin{eqnarray}
 S_{FP}[b,c, v^{++}, V^{++}] = - tr \int d\zeta^{(4)} du
\nabla^{++}b (\nabla^{++}c +
ig[v^{++}, c])
 \end{eqnarray}
$b$ and $c$ are the Faddeev-Popov
ghosts and $f^{(4)}$ is an arbitrary external analytic superfield.

The final step is averaging over all superfields $f^{(4)}$. To do that we multiply both parts of the expression for $e^{i\Gamma_{SYM}}$ with the unit
\begin{eqnarray}
1=\Delta[V^{++}]\int {\cal D}f^{(4)}
exp \{\frac{i}{2 \alpha} tr
\int d^{12}z du_1 du_2 f^{(4)}_{(\tau)}(z, u_1)
\frac{(u^-_1 u^-_2)}{(u^+_1 u^+_2)^3} f^{(4)}_{(\tau)} (z, u_2) \}
\end{eqnarray}
where $\alpha$ is an arbitrary parameter and
\begin{eqnarray}
\Delta [V^{++}]&=& Det^{-\frac{1}{2}}(\nabla^{++})^2 Det^{\frac{1}{2}}
(\stackrel{\frown}{\Box})
\nonumber
\\
\stackrel{\frown}{\Box} &=& -\frac{1}{2}(D^+)^2 (\bar D^+)^2
(\nabla^{--})^2 = \Box + ...
\end{eqnarray}
with $\Box$ be the the
standard dalambertian. The functional determinant
$Det^{-\frac{1}{2}}(\nabla^{++})^2 $ can be
presented by path integral in the form
\begin{eqnarray}
& &Det^{-\frac{1}{2}}(\nabla^{++})^2 = \int {\cal D}\phi e^{iS_{NK}[\phi, V^{++}]}
\nonumber
\\
& &S_{NK}[\phi, V^{++}] = -\frac{1}{2} \int d\zeta^{(-4)} du
tr (\nabla^{++}\phi)
(\nabla^{++}\phi)
\end{eqnarray}
with bosonic real analytic superfield $\phi$. This superfield $\phi$ has sense of so called
Nilsen-Kallosh ghost.

After doing averaging over $f^{(4)}$ and putting $\alpha = -1$ we obtain the final form for effective action
\begin{eqnarray}
e^{i \Gamma_{SYM}[V^{++}]}= e^{i S_{SYM}[V^{++}]} \int {\cal D}v^{++}
{\cal D}b {\cal D}c {\cal D}\phi Det^{\frac{1}{2}}
(\stackrel{\frown}{\Box}) e^{iS_{total}}
\nonumber
\\
S_{total}[v^{++}, b, c, \phi, V^{++}] =
S_2[v^{++}, b, c, \phi, V^{++}] + S_{int}[v^{++}, b, c, V^{++}]
\end{eqnarray}
$S_2$ plays a role of action of free theory
\begin{eqnarray}
& &S_2[v^{++}, b, c, \phi, V^{++}] = - \frac{1}{2} \int d\zeta^{(-4)} du
tr v^{++} \bar{\Box} v^{++} -
\nonumber
\\
 & & - \int d\zeta^{(-4)} du tr (\nabla^{++}b)(\nabla^{++}c) -
\frac{1}{2} \int d\zeta^{(-4)} du tr (\nabla^{++} \phi)(\nabla^{++}\phi)
\nonumber
\\
& &S_{int}[v^{++}, b, c, V^{++}] =
- \int d^{12}z du_1 ... du_n tr \sum^{\infty}_{n=2}
\frac{(-ig)^{n-2}}{n}
\frac{v^{++}_{(\tau)}(z,u_1)v^{++}_{(\tau)}(z, u_2) ...
v^{++}_{(\tau)}(z, u_n)}
{(u^+_1 u^+_2)(u^+_2 u^+_3) ... (u^+_n u^+_1)}
\nonumber
\\
& & - ig \int d\zeta^{(-4)} du tr (\nabla^{++}b)[{v
}^{++}, c]
\end{eqnarray}
The path integral defining $\Gamma_{SYM}[V^{++}]$ has the form standart for quantum field theory. It contains
the free action and interaction. The free action defines the propagators for super
Yang-Mills field and ghosts and the interaction defines the vertices. Obtained form for
effective action opens the possibilities to develop the manifestly $N=2$ supersymmetric
and gauge invariant procedures for calculating the effective action
$\Gamma_{SYM}[V^{++}]$.

\section{General Structure of Effective Action}

As in conventional field theory one can suggest that the effective action $\Gamma [V^{++}]$
for $N=2$ super Yang-Mills theory with matter is described in terms of effective Lagrangians
\begin{eqnarray}
\Gamma [V^{++}] = \int d^4 x d^4 \theta d^4 \bar \theta {\cal L}_{eff} +
( \int d^4 x d^4 \theta  {\cal L}_{eff}^{(c)} + c.c. )
\end{eqnarray}
where ${\cal L}_{eff}$ can be called general effective Lagrangian and
${\cal L}_{eff}^{(c)}$ can be called  chiral effective Lagrangian.

We will assume that the theory under consideration is formulated within background field
method and hence the effective action $\Gamma [V^{++}]$ is gauge invariant under initial classical gauge
transformations. In this case the effective action should be constructed only from strengths
and their covariant derivatives. Therefore  the effective Lagrangians
can be written as follows
\begin{eqnarray}
{\cal L}_{eff} &=& {\cal H}(W, \bar W) + term\; depending\; on\;
covariant\; derivatives\; of\; W\; and\; \bar W
\nonumber
\\
{\cal L}_{eff}^{(c)} &=& {\cal F}(W) + term \; depending\; on\;
covariant\; derivatives\; of\; W\; and\;
\nonumber
\\& & \quad preserving \; chirality
\end{eqnarray}

The term ${\cal F}(W)$ in chiral effective Lagrangian depending only on $W$ is called holomorphic
effective action. Namely this term is leading in low-energy approximation and describes a vacuum
 structure of the theory. The term
\begin{eqnarray}
\int d^4 x d^4 \theta d^4 \bar \theta {\cal H}(W, \bar W)
\end{eqnarray}
defines first non-leading correction to low-energy effective action and describes an effective
dynamics..

The structure of effective action in $N=2$ case turned out to be analogous to structure
effective action depending on chiral and antichiral superfields in $N=1$ case. This low-energy
effective action $\Gamma [\phi, \bar \phi]$ is described by three objects: Kahlerian effective potential $K(\phi, \bar \phi)$,
chiral effective potential $V_{eff}^{(c)}(\phi)$ and so called effective action of auxilliary fields [18,19,16].
The chiral effective potential depends only on $\phi$ and can be called a holomorphic
effective potential. We see that
 $V_{eff}^{(c)}(\phi)$ in $N=1$ case is analogous to holomorphic effective action ${\cal F}(W)$ in
$N=2$ case. The first non-leading correction ${\cal H}(W, \bar W)$ in $N=2$ case is analogous to
Kahlerian effective potential $K(\phi, \bar \phi)$ in $N=1$ case.

\section{One-Loop Holomorphic and  Non-Holomorphic Contributions to Low-Energy
Effective Action for Hypermultiplets}

We define the effective action $\Gamma [V^{++}]$ corresponding to $q$-hypermultiplet coupled to external
super Yang-Mills field by path integral
\begin{eqnarray}
e^{i \Gamma [V^{++}]} = \int {\cal D}\breve{q}^+ {\cal D} q^+
e^{i S[\breve{q}^+ ,q^+ , V^{++}]}
\end{eqnarray}
where
\begin{eqnarray}
S[\breve{q}^+ ,q^+ , V^{++}] = - \int d\zeta^{(-4)} du \breve{q}^+
\nabla^{++} q^+
\end{eqnarray}
One can show  [13] that effective action for
$\omega$-hypermyltiplet $\Gamma_{\omega}[V^{++}]$ can be written as
follows
\begin{eqnarray}
\Gamma_{\omega}[V^{++}] = 2 \Gamma[V^{++}]
\end{eqnarray}
Besides, the one-loop ghosts contribution to effective action within background field
method can also be expressed in terms of  $\Gamma [V^{++}]$  therefore the $q$-hypermultiplet
 in external
gauge superfield is basic model for investigating one-loop effective action.

Formal calculating the above path integral leads to
\begin{eqnarray}
\Gamma [V^{++}] = i Tr ln(\nabla^{++})
\end{eqnarray}
Further we will consider only case of abelian theory.

Let us write again the classical action
\begin{eqnarray}
S[\breve{q}^+ ,q^+ , V^{++}] = - \int d\zeta^{(-4)} du
\breve{q}^+ (D^{++} + iV^{++}) q^+
\end{eqnarray}
It is convenient to decompose the $V^{++}$ into two pieces in the form
\begin{eqnarray}
V^{++} = V^{++}_0 + V^{++}_1
\end{eqnarray}
where $V^{++}_0$ possesses the constant strength and can be expressed as follows [13,20]
\begin{eqnarray}
& &V^{++}_0 = - (\theta^+ )^2 \bar W_0 - (\bar \theta^+ )^2  W_0
\nonumber \\
& &W_0 = const, \quad \bar W_0 = const.
\end{eqnarray}
If $V^{++}_0 = 0$ we have $W_0 = \bar W_0 = 0$ and we get massless hypermultiplet. If $V^{++}_0 \neq 0$ then the equation
of motion
\begin{eqnarray}
(D^{++} + iV^{++}_0)q^+ = 0
\end{eqnarray}
leads to the equation
\begin{eqnarray}
(\Box + m^2)q^+ = 0
\end{eqnarray}
where $m^2=\bar W_0 W_0$. Therefore this case corresponds massive hypermultiplet. The structure
of low-energy effective action depends crucially is $m=0$ or no.

To evaluate the functional trace $Tr ln (V^{++})$ one introduces the Green function
$G^{(1,1)}(1,2)$ of the operator $\nabla^{++}$ by the equation
\begin{eqnarray}
\nabla^{++} G^{(1,1)}(1,2) = \delta_A^{3,1}(1,2)
\end{eqnarray}
where $ \delta_A^{3,1}(1,2)$ is so called analytic $\delta$-function investigated
in the refs. [10,11]. Let us define the  analytic kernel $Q^{(3,1)}(1,2)$ as follows
\begin{eqnarray}
G^{(1,1)}_0 (1,2) = \int d\zeta^{(-4)}_3 du_3 G^{(1,1)}(1,3)
Q^{(3,1)}(3,2)
\end{eqnarray}
where $G^{(1,1)}_0$ is Green function of free hypermultiplet. It can be taken corresponding to
 or massless or  massive theory. The above equation allows to express the kernel
in the form
\begin{eqnarray}
Q^{(3,1)}(1,2) = \delta^{(3,1)}_A (1,2) + i V^{++}(1)G^{(1,1)}_0 (1,2)
\end{eqnarray}
The effective action can be rewritten in terms of $Q^(3,1)$ up to unessential constant
\begin{eqnarray}
\Gamma [V^{++}]= i Tr ln Q^{(3,1)}
\end{eqnarray}
Taking into account the structure of $Q^{(3,1)} \sim 1+iV^{++}G_0$ we see that
$\Gamma [V^{++}]$ is well defined within
perturbation theory
\vspace{3mm}

$\Gamma[V^{++}]=\sum\limits_{n=1}^\infty\Gamma_n[V^{++}]={\rm i}^2$\
\wavelin(1,0)\mov(1,0)\vdot\mov(1.35,0){\Circle(0.7)}\mov(2,0)
{$-\,\displaystyle\frac{1}{2}{\rm i}^3$}
\mov(2.7,0){\wavelin(1,0)}\mov(3.7,0)\vdot\mov(4.05,0){\Circle(0.7)}
\mov(4.3,0)\vdot\mov(4.3,0){\wavelin(1,0)}\mov(5.7,0){+}

$+\,\displaystyle\frac{1}{3}{\rm i}^4$\
\wavelin(1,0)\mov(1,0)\vdot\mov(1.35,0){\Circle(0.7)}\mov(1.7,0)
{\wavelin(1,0)}\mov(1.7,0)\vdot\mov(1.35,0.35)\vdot
\mov(1.35,0.35){\wavelin(0,0.7)}\mov(3,0)
{$+\ \dots\ +\,\displaystyle\frac{(-1)^{n+1}}{n}(-{\rm i})^{n+1}$}
\mov(6.5,0){\Circle(0.7)}\mov(6.2,-0.2)\vdot\mov(6.8,-0.2)\vdot
\mov(6.2,0.2)\vdot\mov(6.8,0.2)\vdot\mov(6.5,-0.35)\vdot
\mov(6.5,-0.35){\wavelin(0,-0.7)}
\mov(6.15,-0.2){\wavelin-(-0.5,-0.5)}
\mov(6.65,-0.2){\wavelin(0.5,-0.5)}
\mov(5.95,0.2){\wavelin-(-0.5,0.5)}
\mov(6.45,0.2){\wavelin(0.5,0.5)}
\mov(5.75,0.5)\vdot\mov(6.05,0.6)\vdot\mov(6.35,0.5)\vdot
\mov(7,0){$+\ \dots$}

\vspace{3mm}

\noindent
Here the wave line corresponds to external field $V^{++}$, the term $\Gamma_n [V^{++}]$
is given by supergraph with n external lines.

Let as discuss briefly the results of supergraphs calculations leading to
holomorphic and non-holomorphic contributions to effective action [13].

\noindent
(i) Massless theory. Taking into account the structure of free Green function
$G_0^{(1,1)}$ one can show that $\Gamma_n [V^{++}] =0$ at odd n and holomorphic
contribution is absent at all.
The first non-holomorphic contribution is $\Gamma_4 [V^{++}]$ and corresponds to four-leg
supergraph. The straightforward calculation [13] lead to
\begin{eqnarray}
\Gamma_4 [V^{++}] = \frac{1}{(16\pi)^2 \Lambda^4}
\int d^4 x d^8 \theta \bar W^2 W^2
\end{eqnarray}
where $\Lambda$ is infrared cutoff. This result has a simple physical interpretation.
Let us keep non-vanishing only electromagnetic field component of $W$, that is
$F_{\mu\nu}$. Then
\begin{eqnarray}
\Gamma_4 [V^{++}] = \frac{1}{(16\pi)^2 \Lambda^4}
\int d^4 x [
(F_{\mu\nu}F^{\mu\nu})^2 + (F_{\mu\nu}\tilde F^{\mu\nu})^2
\end{eqnarray}
where $\tilde F_{\mu\nu}$ is dual to $F_{\mu\nu}$. This type of nonlinear corrections to
Maxwell Lagrangian was originally discovered by Heisenberg and Euler (see f.e. [21]).
 Therefore our $\Gamma_4 [V^{++}]$ can be interpreted
as the $N=2$ supersymmetric generation of Heisenberg-Euler effective Lagrangian.

\noindent
(ii)Massive theory. In this case $V^{++}_0 \neq 0$. We can act  here by two ways.
\begin{enumerate}
\item
The free propagator $G_0^{(1,1)}$ corresponds to massless theory and we consider
$V^{++} = V^{++}_0 + V^{++}_1$ for external lines in the supergraphs.
\item
The free propagator $G_0^{(1,1)}$ corresponds to massive theory and we consider only
$V^{++}_1$ for external lines.
\end{enumerate}

Both these ways lead to the same results. Holomorphic effective action is
obtained in the form
\begin{eqnarray}
{\cal F}(W)= \frac{1}{64 \pi^2}W^2 (1- ln\frac{W^2}{\mu^2})
\end{eqnarray}
where $\mu$ is renormalizaton scale. Imposing the renormalization condition
\begin{eqnarray}
{\cal F}(W)_{| W^2=M^2} = 0
\end{eqnarray}
with some scale $M$ one gets finally
\begin{eqnarray}
{\cal F}(W)= - \frac{1}{64 \pi^2}W^2  ln\frac{W^2}{M^2}
\end{eqnarray}
This function coincides on its structure with holomorphic effective action
obtained in ref. [9]. Of course, we have another coefficient since we considere
another theory.

In massive case we also will have non-holomorphic contribution which is not
written here.

\section{Summary and Open Problems}

Let us summarize the results

\noindent
(i) The background field method for $N=2$ super Yang-Mills theory has developed within
harmonic superspace approach. Ghost structure of the theory is established. Path integral for manifestly $N=2$
supersymmetric and gauge invariant effective action is presented. This path integral has standarted quantum
field theory form where quadratic part of total action defines the propagators and all other parts defines the vertices.

\noindent
(ii) The general approach to effective action of $N=2$ abelian superfield coupled to hypermultiplet has developed.
This approach is based on formulation of $N=2$ theories in harmonic superspace and garantees manifest $N=2$
supersymmetry on all stages of calculations. The approach under consideration allows to calculate straightforwardly
both holomorphic and non-holomorphic contributions to low-energy effective action.

In conclusion we should like to point out some open problems associated with effective
action of $N=2$ supersymmetric field theories in harmonic superspace approach.

\noindent
(i) Developing a proper-time technique providing a most efficient way of gauge-invariant calculations of
effective action.

\noindent
(ii) Calculation of terms in effective action depending on the derivatives of
 $W$ and $\bar W$.

\noindent
(iii) Higher-loop contributions to effective action.

\noindent
(iiii) Quantum aspects of $N=2$ supergravity in harmonic superspace approach.

\vspace{0.7cm}

{\bf Acknowledgements}

I am very grateful to my co-authors E.I. Buchbinder, E.A. Ivanov, S.M. Kuzenko and
B.A. Ovrut for collaboration and valuable discussions. The work was partially supported
by the grants of RFBR, project No 96-02-16017; RFBR-DFG, project No 96-02-00180,
by INTAS grant, INTAS 96-0308. I am very indebted to Department of Physics, University
of Pennsylvania where part of this research was carried out, and for support from the
Research Foundation of the  University of Pennsylvania.

\newpage

{\large{\bf References}}
\begin{enumerate}

\item
P. Fre, P. Soriani, The $N=2$ Wonderland, Word Scientific, Singapore, 1995
\item
S.J. Gates, M.T. Grisaru, M. Ro\u{c}ek, W. Siegel, Superspace, Benjamin-Cummings,
Reading, MA, 1983.
\item
P. Howe, K. Stelle, P. West, Phys. Lett. B124 (1983) 55
\item
P.West, Supersymmetry and Finiteness, in Proceeding of the 1983 Shelter Island II Conference
on  Quantum Field Theory and Fundamental Problems of Physics, edited by R. Jakiw,, N. Kuri,
S. Weinberg and E. Witten, M.I.T. Press, 1983
\item
P.S. Howe, K. Stelle, P.K. Townsend, Nucl. Phys. B236 (1984) 125
\item
P. West, Introduction to Supersymmetry and Supergravity, World Scietific, Singapore, 1990
\item
I.L. Buchbinder, S.M. Kuzenko, B.A. Ovrut,  On the D=4, N=2 non-renormalization theorem, hep-th/9710142
\item
N. Seiberg, E. Witten, Nucl. Phys. B426 (1994) 19; B430 (1994) 485
\item
N. Seiberg, Phys. Lett. B206 (1988) 75
\item
A. Galperin, E. Ivanov, S. Kalitzin, V. Ogievetsky, E. Sokatchev, Class. Quant. Grav. 1(1984) 469
\item
A. Galperin, E. Ivanov, V. Ogievetsky, E. Sokatchev, Class. Quant. Grav. 2(1985) 601; 617
\item
B. de Wit, M.T. Grisaru, M. Ro\u{c}ek, Phys. Lett. B374 (1996) 297;
A. Pickering, P. West, Phys. Lett. B383 (1996) 54;
M.T. Grisaru, M. Ro\u{c}ek, R. van Unge, Phys. Lett. B383 (1996) 415;
T.E. Clark, S.T. Love, Phys. Lett. B388 (1996) 577;
U. Lindst\"{o}m, F. Gonzales-Rey, M. Ro\u{c}ek, R. van Unge, Phys. Lett. B388 (1996) 581;
A. De Giovanni, M.T. Grisaru, M.Ro\u{c}ek, R. van Unge, D. Zanon, Phys. Lett. B409 (1997) 251;
A. Yung, Nucl. Phys. B 485 (1997) 38;
M. Matone, Phys. Rev. Lett. 78 (1997) 1412;
D. Bellisai, F. Fucito, M. Matone, G. Travaglini, Phys. Rev. D56 (1997) 5218.
\item
I.L. Buchbinder, E.I. Buchbinder, E.A. Ivanov, S.M. Kuzenko, B.A. Ovrut, Phys. Lett. B412 (1997) 309.
\item
I.L. Buchbinder, E.I. Buchbinder, S.M. Kuzenko, B.A. Ovrut,
Phys. Lett. B417 (1998) 61.
\item
B. Zupnik, Teor. Mat. Fis. (Theoretical and Mathematical Physics, in Russian) 69 (1986) 207;
Phys. Lett. B183 (1987) 175.
\item
I.L. Buchbinder, S.M. Kuzenko, Ideas and Methods of Supersymmetry and Supergravity,
IOP Publ, Bristol and Philadelphia, 1995
\item
R. Grimm, M. Sohnius, J. Wess, Nucl. Phys. B133 (1978) 275
\item
I.L. Buchbinder, S.M. Kuzenko, J. V. Yarevskaya, Yad. Fiz. (Physics of Atomic Nuclei, in Russian)
56 (1993) 202; Nucl. Phys. B411 (1994) 665.
\item
I.L. Buchbinder, S.M. Kuzenko, A. Yu. Petrov, Phys. Lett. B321 (1994) 372;
Yad. Fiz. (Physics of Atomic Nuclei, in Russian) 59 (1996)157
\item
I.L. Buchbinder, S.M. Kuzenko, Class. Quan. Grav. 14 (1997) L157
\item
C. Itzykson, J. -B. Zuber, Quantum Field Theory, Mc Graw - Hill Book Company, Ney-York, 1980

\end{enumerate}

\end{document}